# Investigation of Heat Transfer Rates on Shell and Tube Heat Exchanger by Numerical Modelling via CFD Analysis

*HRUTUJ RAUT, DEPARTMENT OF CHEMICAL ENGINEERING, VISHWAKARMA INSTITUTE OF TECHNOLOGY, PUNE, INDIA*

**ABSTRACT**

*A shell and tube heat exchanger design with respect to the total heat transfer rate and temperature profile has been investigated by numerical modelling. The HE comprises of a single tube with a length of 200mm and a shell diameter of 20mm and has been resolved by the two equation models namely, the k – epsilon and k – omega. The low RMS residual levels calculated showed that the convergence is better in the k-ω model. The differences in the heat transfer rate noted are likely due to the low number of iterations in the k-ε model and a loose convergence. The study of the effect of residuals and iterations on the flow and heat transfer helps explain the selection of an appropriate model - k-ω. In the operating conditions used for the simulation, it is observed that k-ω model has better heat transfer rate. It is arguable that the heat transfer was not accurate in the k-ε model owing to a lesser number of iterations leading to a higher residual level. Hence, it is safe to conclude that k-ω model is best suited for the heat exchanger.*

**Keywords** – *Heat Exchanger, CFD, k – epsilon, k – omega*

# TABLE OF CONTENTS

# LIST OF TABLES

# LIST OF EQUATIONS

# LIST OF FIGURES

# 1 Introduction

## 1.1 Background

The general function of a heat exchanger is to transfer heat from one fluid to another. The basic component of a heat exchanger can be viewed as a tube through which one fluid passes and the other fluid flows out. Heat exchangers are usually classified by flow arrangement and construction type. The simplest heat exchanger is one in which the hot and cold fluids move in the same or opposite direction in a concentric tube (or dual tube) design. In a counter-current arrangement, fluids enter at opposite ends, flow in opposite directions, and exit at opposite ends. The performance and efficiency of heat exchangers is measured by the amount of heat transferred using the smallest heat transfer area and pressure drop. A better presentation of its effectiveness is done by calculating over all heat transfer coefficients. The pressure-drop and area required for a given amount of heat transfer provides an overview of the capital costs and energy requirements (operating costs) of the heat exchanger. There is usually a lot of literature and theory on how to design a heat exchanger according to requirements. A good design means a heat exchanger with the smallest possible surface area and pressure drop to meet the heat transfer requirements. (1)

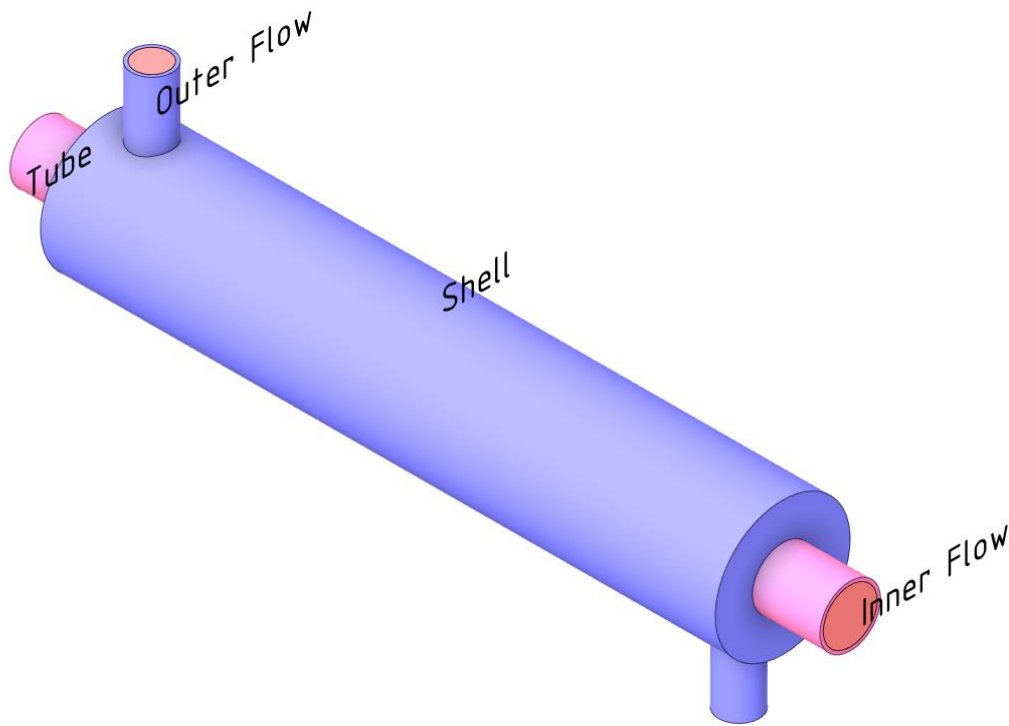

*Figure 1: Heat Exchanger*

## 1.2 Heat Exchanger Classification

In a shell and tube heat exchanger, the coolant usually flows through a central "tube core" to cool the hot oil, water, or air that passes through and around the tubes. The direction in which the two fluids pass through the heat exchanger can be either "parallel flow" or "counter-current". Parallel flow is where the fluid to be cooled flows through the heat exchanger in the same direction as the cooling medium. In counter-current cooling, the inlet cooling medium absorbs heat while the "hot" fluid moves in the opposite direction. The coolant heats up as it passes through the heat exchanger, but when cooler water enters the heat exchanger, it absorbs more heat and lowers the temperature much lower than could be achieved with parallel flow. (2)

While a heat exchanger installed with parallel flow will reduce the temperature, it is not nearly as efficient as a counter-flow arrangement, and a larger heat exchanger may eventually be required to achieve the desired outlet temperature. In contrast, counterflow is significantly more efficient and, depending on flow and temperature, heat transfer performance can be up to 15% more efficient, potentially allowing the use of a smaller heat exchanger, saving space and money. (2)

## 1.3 APPLICATIONS OF HEAT EXCHANGERS

The application of heat exchangers is a very vast subject and would require a separate thorough study to cover every aspect. Common applications include their use in the manufacturing, machinery and home appliance industries. Heat exchangers can be found used for district heating systems which are widely used nowadays. Air conditioners and refrigerators also install heat exchangers to condense or evaporate the liquid. In addition, they are also used in milk processing units for pasteurization. (3)

| **Industries** | **Applications** |
|---|---|
| Food and Beverages | Ovens, cookers, Food processing and pre-heating, Milk pasteurization, beer cooling and pasteurization, juices and syrup pasteurization, cooling or chilling the final product to desired temperatures. |
| Petroleum | Brine cooling, crude oil pre-heating, crude oil heat treatment, fluid interchanger cooling, acid gas condenser. |
| Hydrocarbon processing | Preheating of methanol, liquid hydrocarbon product cooling, feed pre-heaters, recovery or removal of carbon dioxide, production of ammonia. |
| Polymer | Production of polypropylene, reactor jacket cooling for the production of polyvinyl chloride. |
| Pharmaceutical | Purification of water and steam, for point of use cooling on water for injection ring. |
| Automotive | Pickling, Rinsing, Priming, Painting. |
| Power | Cooling circuit, Radiators, Oil coolers, air conditioners and heaters, energy recovery. |
| Marine | Marine cooling systems, Fresh water distiller, Diesel fuel pre-heating, central cooling, Cooling of lubrication oil. |

*Table 1: Application of HE*

## 1.4 HEAT EXCHANGER

### 1.4.1 Heat Transfer

Heat transfer is considered a fundamental process of all manufacturing industries. During the heat transfer process, one fluid at a higher temperature transfers its energy in the form of heat to another fluid at a lower temperature. A fluid can transfer its heat by various mechanisms. These heat transfer mechanisms are conduction, convection, and radiation. Radiation is not a common method of heat transfer in the manufacturing industry, but in some processes, it plays a vital role in heat transfer, for example in an incinerator. The other two methods of heat transfer, i.e., conduction and convection, are the most common methods of heat transfer in the manufacturing industry. (4) (5)

Overall Energy Balance of a heat transfer system can be written by using the Equation 1 and Equation 2.

$$Q_h = m_h C_p (T_{h,i} - T_{h,o})$$

*Equation 1: Energy balance of heat transfer system*

$$Q_c = m_c C_p (T_{c,o} - T_{c,i})$$

*Equation 2: Energy balance of heat transfer system*

In reality, the heat provided by the hotter fluid to the cold fluid is not exactly the same due to losses and resistances in the form of wall fouling. It is assumed that the amount of heat transferred from the warmer fluid is equal to the amount of heat transferred to the cooler fluid. Usually, heat exchangers are insulated to minimize losses to the environment.

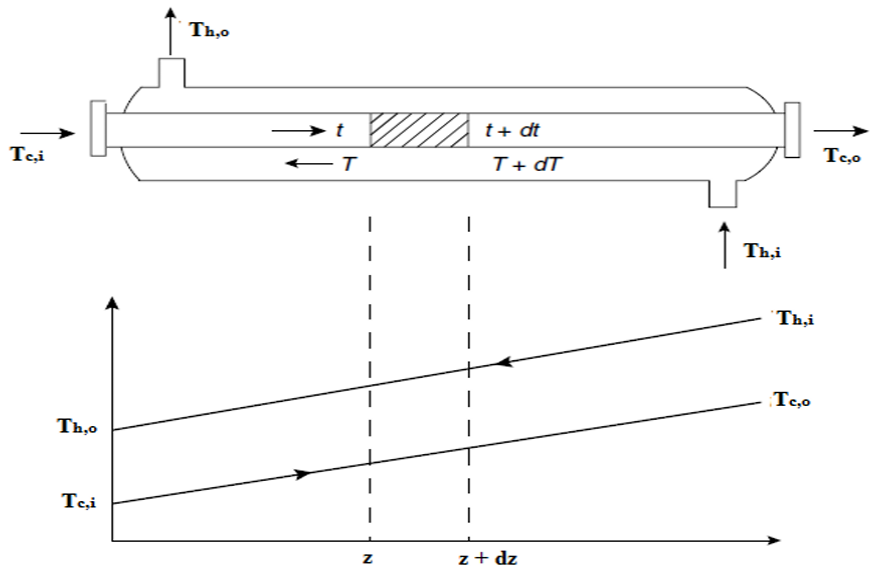

*Figure 2: Heat Transfer in a Heat Exchanger (6)*

$$Q_h = Q_c = Q$$

Graphical representation of these equations makes the process easier and simpler to understand. These graphs are known as T-Q diagrams. These graphs also help ensure that the 2nd law of thermodynamics is obeyed, i.e., heat should always be transferred from a higher temperature to a lower temperature. Then we can write the equation as,

$$Q = UA\Delta T_{LM}$$

*Equation 3: Equation of Heat Transfer*

Where,

Q = Heat transfer rate (W)

A = Heat transfer area (m²)

U = Overall heat transfer coefficient (W/m².K)

$\Delta T_{LM}$ = Logarithmic mean temperature difference (K)

These three equations: Equation 1, Equation 2 and Equation 3 are considered the fundamental equations for all heat transfer problems.

These equations are derived using various assumptions. Above all, the overall heat transfer coefficient and the specific heat capacity are considered constant for heat exchangers. In real practice, these values may change depending on the properties of liquids and temperatures.

The specific heat of a fluid is the property of the fluid that transfers heat. In other words, it is the amount of heat that one kilogram of liquid needs to raise its temperature by one degree Celsius. The log mean temperature difference (LMTD) is calculated to estimate the average temperature difference across the heat exchanger. It is basically the logarithmic average of the temperature difference. For heat transfer, the driving force is always the temperature difference, so a higher log-mean temperature difference will ensure better heat transfer. This is related to the heat exchanger area such that a higher LMTD will result in a smaller heat transfer area and a lower LMTD will need a larger heat transfer area. In general, LMTD is a process condition and not much can be done about it because the inlet and outlet fluid temperatures are usually predetermined for the design of the heat exchanger. The area can certainly be reduced by making full use of the available LMTDs by efficient heat transfer. (7)

### 1.4.2 Heat Transfer Coefficient

The total heat transfer of heat exchangers is the ability to transfer heat through different resistances, it depends on the properties of the process fluids, temperatures, flow rates and the geometric arrangement of the heat exchanger. For example, the number of passages, the number of baffles and the spacing of the baffles, etc. It is defined by the Equation 4. This equation basically sums up all the resistances encountered in heat transfer and flipping it gives us the overall heat transfer coefficient. (5)

$$\frac{1}{U} = \frac{1}{h_h} + \frac{\Delta x}{k} + \frac{1}{h_c} + R_f$$

*Equation 4: Heat Transfer Coefficient*

Where,

$h_h$ = Hot side heat transfer coefficient (W/m².K)

$h_c$ = Cold side heat transfer coefficient (W/m².K)

$\Delta x$ = Exchanger tube wall thickness (m)

k = Exchanger wall material thermal conductivity (W/m.K)

$R_f$ = Fouling coefficient (W/m².K)

The equation for the overall heat transfer coefficient can be written as Equation 5.

$$\frac{1}{U} = \frac{1}{h_h} + \frac{1}{h_c} + R_f$$

*Equation 5: Overall Heat Transfer Coefficient*

$h_h$ and $h_c$ are individual film coefficients and are defined as heat transfer rates per unit area and unit temperature difference. These are calculated separately for external and internal liquids. The temperature difference between the average bulk fluid temperature (hot and cold) and the wall temperature (inside and outside) is the driving force for the respective fluids. $\Delta x/k$ is usually ignored because it has no significant effect on the overall heat transfer coefficient. (5)

# 2 MATHEMATICAL BACKGROUND

This chapter explains the governing equations solved with FLUENT and the turbulence models used for this simulation. Two equation models are used for the simulations. The flow equations and energy equations are also described in detail.

## 2.1 FLOW CALCULATIONS

The flow is governed by the continuity equation, the energy equation, and the Navier-Stokes momentum equations. Mass, energy and momentum are transferred by convective flow and diffusion of molecules and turbulent eddies. (8)

### 2.1.1 Continuity Equation

This equation describes the conservation of mass and is written as,

$$\frac{\partial \rho}{\partial t} + \frac{\partial \rho U_1}{\partial x_1} + \frac{\partial \rho U_2}{\partial x_2} + \frac{\partial \rho U_3}{\partial x_3} = 0$$

*Equation 6: Continuity Equation*

Defines the rate of increase of mass in a control volume as an equal amount over its area.

### 2.1.2 Navier-Stokes Equation

Momentum balance, also known as the Navier-Stokes equations, obeys Newton's second law: The change in momentum in all directions is equal to the sum of the forces acting in those directions. Two different kinds of forces act on a finite volume element, surface forces and body forces. Surface forces include pressure and viscous forces, and body forces include gravitational, centrifugal, and electromagnetic forces. (8)

This equation can be written as (for a Newtonian fluid),

$$\frac{\partial U_i}{\partial t} + U_j \frac{\partial U_i}{\partial x_j} = -\frac{1}{\rho}\frac{\partial \rho}{\partial x_i} + \nu \frac{\partial}{\partial x_j}\left(\frac{\partial U_i}{\partial x_j} + \frac{\partial U_j}{\partial x_i}\right) + g_i$$

*Equation 7: Navier-Stokes Equation*

In addition to gravity, there may be other external sources that can affect the acceleration of the fluid, such as electric and magnetic fields. Strictly speaking, it is the momentum equations that make up the Navier-Stokes equations, but sometimes the continuity and momentum equations together are called the Navier-Stokes equations. The Navier-Stokes equations are limited to macroscopic conditions. (8)

The continuity equation is difficult to solve numerically. In CFD programs, the continuity equation is often combined with the momentum equation to form Poisson's equation. For constant density and viscosity, the equation can be written as,

$$\frac{\partial}{\partial x_i}\left(\frac{\partial P}{\partial x_i}\right) = -\frac{\partial}{\partial x_i}\left(\frac{\partial (\rho U_i U_j)}{\partial x_j}\right)$$

*Equation 8: Poisson's Equation*

This equation has more suitable numerical properties and can be solved by suitable iterative methods.

### 2.1.3 Energy Equation

Energy is present in a flow in many forms, i.e., as kinetic energy due to the mass and velocity of the fluid, as thermal energy and as chemically bound energy. So, the total energy can be defined as the sum of all these energies. (8)

$$h = h_m + h_T + h_C + \Phi$$

*Equation 9: Energy Equation*

| | |
|---|---|
| $h_m = \dfrac{1}{2}\rho U_i U_i$ | Kinetic Energy |
| $h_T = \sum_n m_n \int_{T_{ref}}^{T} C_{p,n}\, dT$ | Thermal Energy |
| $h_C = \sum_n m_n h_n$ | Chemical Energy |
| $\Phi = g_i x_i$ | Potential Energy |

*Table 2: General energy equations*

In the above equations, $m_n$ and $C_{p,n}$ are the mass fraction and specific heat for species n. The transport equation for the total energy can be written using the above equations. The coupling between the energy and momentum equations is very weak for incompressible flows, so the kinetic and thermal energy equations can be written separately. Chemical energy is not included as no mode of transport was involved in this project.

### 2.1.4 Two – Equation Models

Different turbulence models can be classified based on the number of additional equations used to close the set of equations. There are zero-, one-, and two-equation models commonly used to model turbulence. The null equation model makes a simple assumption of constant viscosity (Prandtl mixing length model). While one equation model assumes that the viscosity is related to the historical effects of turbulence in relation to the time-averaged kinetic energy. Similarly, a two-equation model uses two equations to close a set of equations. These two equations can model turbulent velocity or turbulent length scales. There are many variables that can be modelled, such as vorticity scale, frequency scale, time scale, and dissipation rate. Of these variables, the most commonly used variable is the ε dissipation rate. This model is named with respect to the modelled variables. For example, the k-ε model because it models k (turbulent kinetic energy) and k-ε (turbulent energy dissipation rate). Another important turbulence model is the k-ω model. It models k (Turbulent Kinetic Energy) and ω (Specific Dissipation Rate). These models have now become common in industrial use. These provide a significant amount of confidence because they use two variables to close the set of equations. (9) (10)

#### 2.1.4.1 k-epsilon Model

k-epsilon Turbulence model, also known as the k-epsilon model was proposed for turbulent bounded flows. The model is divided into two equations. The first equation gives the first transported variable kinetic energy k. The equation is given as,

$$\frac{\partial(\rho k)}{\partial t} + \frac{\partial(\rho k u_i)}{\partial x_i} = \frac{\partial}{\partial x_j}\left[\frac{\mu_t}{\sigma_k}\frac{\partial k}{\partial x_j}\right] + 2\mu_t E_{ij}E_{ij} - \rho\varepsilon$$

*Equation 10: K-epsilon model*

The second equation gives the rate of dissipation of turbulent kinetic energy ε, which is the second transported variable. The equation is given as,

$$\frac{\partial(\rho\varepsilon)}{\partial t} + \frac{\partial(\rho\varepsilon u_i)}{\partial x_i} = \frac{\partial}{\partial x_j}\left[\frac{\mu_t}{\sigma_\varepsilon}\frac{\partial \varepsilon}{\partial x_j}\right] + C_{1\varepsilon}\frac{\varepsilon}{k}2\mu_t E_{ij}E_{ij} - C_{2\varepsilon}\rho\frac{\varepsilon^2}{k}$$

*Equation 11: K-epsilon model*

### 2.1.4.2 K-omega Model

A two-equation model for the prediction of turbulence kinetic energy k and specific rate of dissipation ω. The two equations are given as,

$$\frac{\partial(\rho k)}{\partial t} + \frac{\partial(\rho u_j k)}{\partial x_j} = \rho P - \beta^* \rho \omega k + \frac{\partial}{\partial x_j}\left[\left(\mu + \sigma_k \frac{\rho k}{\omega}\right)\frac{\partial k}{\partial x_j}\right], \quad with\ P = \tau_{ij}\frac{\partial u_i}{\partial x_j}$$

*Equation 12: K-omega model*

$$\frac{\partial(\rho \omega)}{\partial t} + \frac{\partial(\rho u_j \omega)}{\partial x_j} = \frac{\alpha \omega}{k}\rho P - \beta \rho \omega^2 + \frac{\partial}{\partial x_j}\left[\left(\mu + \sigma_\omega \frac{\rho k}{\omega}\right)\frac{\partial \omega}{\partial x_j}\right] + \frac{\rho \sigma_d}{\omega}\frac{\partial k}{\partial x_j}\frac{\partial \omega}{\partial x_j}$$

*Equation 13: K-omega model*

The k-omega model is specifically used for the prediction of boundary-layer transition and flows with low Reynolds number.

SST models stand for Shear Stress Transport model, it is a combination of models k-omega and k-epsilon. Since k-epsilon is a high Reynolds number model, model k-omega is used in the near wall region. While in the area away from the walls, the k-epsilon model is used. The SST model uses a blending function whose value depends on the distance from the walls. At the wall, in the viscous sublayer, this blending function is used only as the k-omega model. Areas away from the wall have this function zero and only use the k-epsilon model. This model also includes a cross-diffusion term. In this model, the turbulent viscosity is modified to include the effect of turbulent shear stress transfer. Modelling constants are also different from other models. These properties make the SST model reliable for adverse pressure gradient flow and boundary layer separation. (11)

## 3 CFD ANALYSIS

### 3.1 GEOMETRY

The geometry of the heat exchanger is built into the design module of ANSYS workbench. The geometry is simplified to account for planar symmetry and is cut in half vertically. It is a counter current HE with a single. The jacket outlet length is also increased to facilitate the modelling program to avoid the backflow condition.

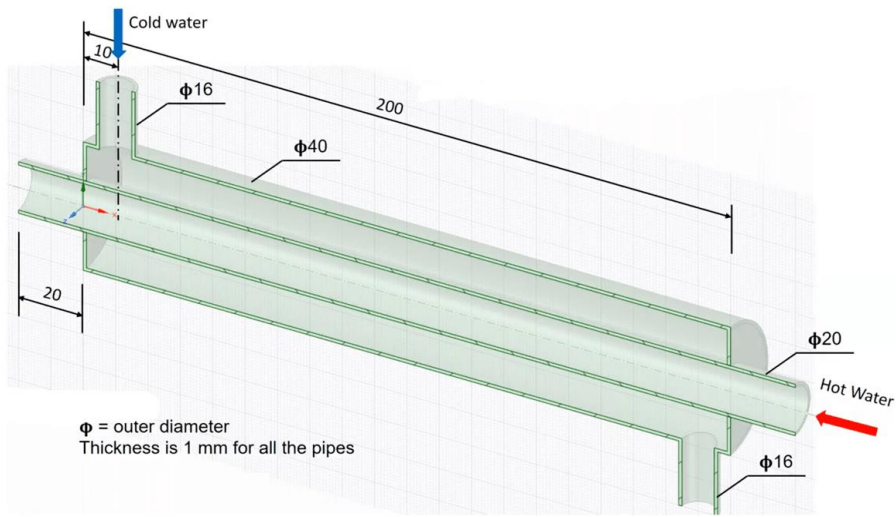

*Figure 3: Geometry of HE*

| Object Name | Geometry |
|---|---|
| State | Fully Defined |
| *Definition* | |
| Type | SpaceClaim |
| Length Unit | Meters |
| *Bounding Box* | |
| Length X | 0.24 m |
| Length Y | 8.e-002 m |
| Length Z | 4.e-002 m |
| *Properties* | |
| Volume | 2.6608e-004 m³ |
| Scale Factor Value | 1 |
| *Statistics* | |
| Bodies | 4 |
| Active Bodies | 4 |

*Table 3: Geometry*

| Object Name | Tube | Shell | Outer_flow | Inner_flow |
|---|---|---|---|---|
| State | Meshed | | | |
| *Definition* | | | | |
| Coordinate System | Default Coordinate System | | | |
| Treatment | None | | | |
| Reference Frame | Lagrangian | | | |
| *Bounding Box* | | | | |
| Length X | 0.24 m | 0.2 m | 0.198 m | 0.24 m |
| Length Y | 1.8e-002 m | 8.e-002 m | | 1.6e-002 m |
| Length Z | 1.8e-002 m | 4.e-002 m | 3.8e-002 m | 1.6e-002 m |
| *Properties* | | | | |
| Volume | 1.2817e-005 m³ | 2.7516e-005 m³ | 1.7749e-004 m³ | 4.8255e-005 m³ |
| Centroid X | 1.e-001 m | 0.1 m | 1.e-001 m | |
| Centroid Y | -9.1116e-019 m | 1.1404e-006 m | -1.7747e-007 m | -1.4911e-019 m |
| Centroid Z | -4.4555e-019 m | -4.7408e-007 m | 7.3789e-008 m | -4.6961e-020 m |

*Table 4: Geometry (parts)*

## 3.2 MESH

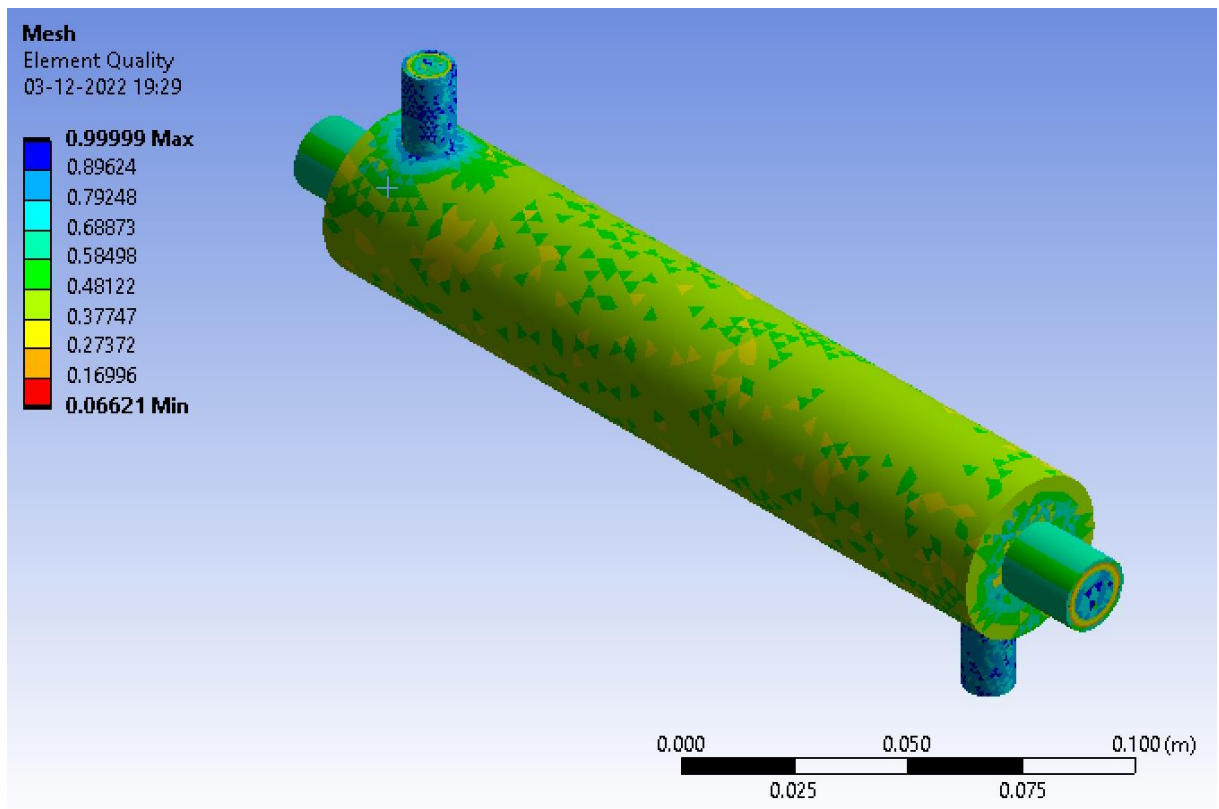

*Figure 4: Mesh of HE (Quality)*

| Object Name | Mesh |
|---|---|
| *State* | Solved |
| *Display* | |
| *Display Style* | Use Geometry Setting |
| *Defaults* | |
| *Physics Preference* | CFD |
| *Solver Preference* | Fluent |
| *Element Order* | Linear |
| *Element Size* | Default (1.2806e-002 m) |
| *Export Format* | Standard |
| *Sizing* | |
| *Use Adaptive Sizing* | No |
| *Growth Rate* | Default (1.2) |
| *Max Size* | Default (2.5612e-002 m) |
| *Mesh Defeaturing* | Yes |
| *Defeature Size* | Default (6.4031e-005 m) |
| *Capture Curvature* | Yes |
| *Curvature Min Size* | Default (1.2806e-004 m) |
| *Curvature Normal Angle* | Default (18.0°) |
| *Capture Proximity* | No |
| *Bounding Box Diagonal* | 0.25612 m |
| *Average Surface Area* | 3.9681e-003 m² |
| *Minimum Edge Length* | 3.1416e-002 m |
| *Quality* | |
| *Target Skewness* | Default (0.900000) |
| *Smoothing* | Medium |
| *Inflation* | |
| *Use Automatic Inflation* | None |
| *Inflation Option* | Smooth Transition |
| *Transition Ratio* | 0.272 |
| *Maximum Layers* | 5 |
| *Growth Rate* | 1.2 |
| *Statistics* | |
| *Nodes* | 54600 |
| *Elements* | 143963 |

*Table 5: Mesh*

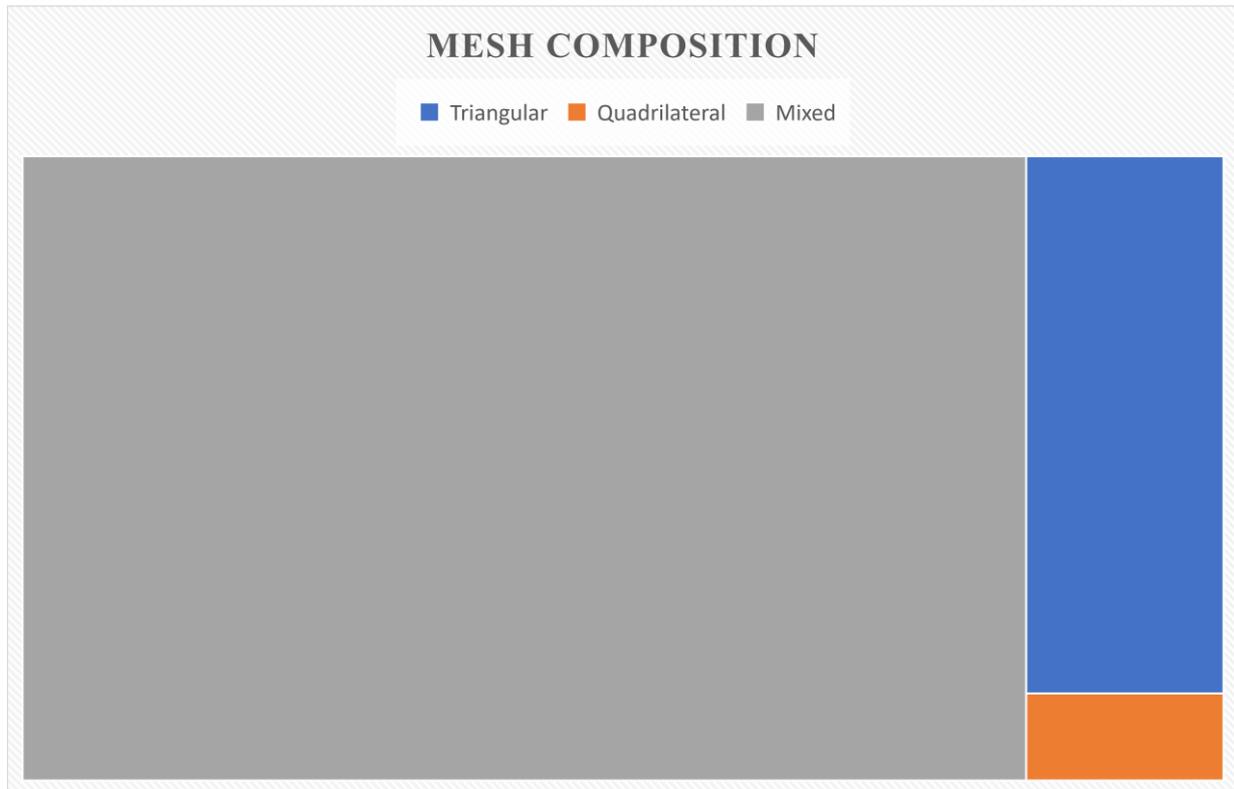

*Figure 5: Mesh Composition (Tree-map representation)*

## 3.3 MATERIALS USED

### 3.3.1 Fluid

#### 3.3.1.1 Water-liquid

The heat transfer fluids transfer the heat to the storage tank and then to the steam generator. Consequently, it is important that good fluids have low viscosity and high heat capacity. Water, synthetic oil and molten salt can be used as heat transfer fluids. Water is a good heat transfer medium because it has a high heat capacity and low viscosity. Its use is cheaper because its use in direct steam production saves costs in the heat exchanger.

|  |  |
|---|---|
| *Parameters* | |
| *Density (kg/m³)* | 998.2 |
| $C_p$ *(Specific Heat) (k/kg-k)* | 4182 |
| *Thermal Conductivity (w/m-k)* | 0.6 |
| *Viscosity* | 0.001003 |

*Table 6: Parameters (Water)*

### 3.3.2 Solid

#### *3.3.2.1 Aluminium*

| *Parameters* | |
|---|---|
| *Density (kg/m³)* | 2719 |
| *$C_p$ (Specific Heat) (k/kg-k)* | 871 |
| *Thermal Conductivity (w/m-k)* | 202.4 |

*Table 7: Parameters (Aluminium)*

#### *3.3.2.2 Copper*

Copper has many desirable properties for thermally efficient and durable heat exchangers. First and foremost, copper is an excellent conductor of heat. This means that the high thermal conductivity of copper allows heat to pass quickly. Other desirable properties of copper in heat exchangers include its corrosion resistance, biofouling resistance, maximum allowable stress and internal pressure, yield strength, fatigue strength, hardness, thermal expansion, specific heat, antimicrobial properties, tensile strength, limit flow, high melting point, alloy, ease of manufacture and ease of joining.

| *Parameters* | |
|---|---|
| *Density (kg/m³)* | 8978 |
| *$C_p$ (Specific Heat) (k/kg-k)* | 381 |
| *Thermal Conductivity (w/m-k)* | 387.6 |

*Table 8: Parameters (Copper)*

## 3.4 BOUNDARY CONDITIONS

### 3.4.1 Inlet

#### *3.4.1.1 inlet_inner_tube*

| *Velocity Magnitude (m/s)* | 0.01 |
|---|---|
| *Temperature (Celsius)* | 30 |

*Table 9: Boundary Conditions (inlet inner tube)*

#### *3.4.1.2 inlet_outer_tube*

| *Velocity Magnitude (m/s)* | 0.05 |
|---|---|
| *Temperature (Celsius)* | 25 |

*Table 10: Boundary Conditions (inlet outer tube)*

### 3.4.2 Outlet

#### *3.4.2.1 outlet_inner_tube*

| *Type* | Pressure Outlet |
|---|---|
| *Backflow Total Temperature (Celsius)* | 26.85 |

*Table 11: Boundary Conditions (outlet inner tube)*

#### *3.4.2.2 outlet_outer_tube*

| *Type* | Pressure Outlet |
|---|---|
| *Backflow Total Temperature (Celsius)* | 26.85 |

*Table 12: Boundary Conditions (outlet outer tube)*

# 4 RESULTS AND ANALYSIS

## 4.1 SCALED RESIDUALS

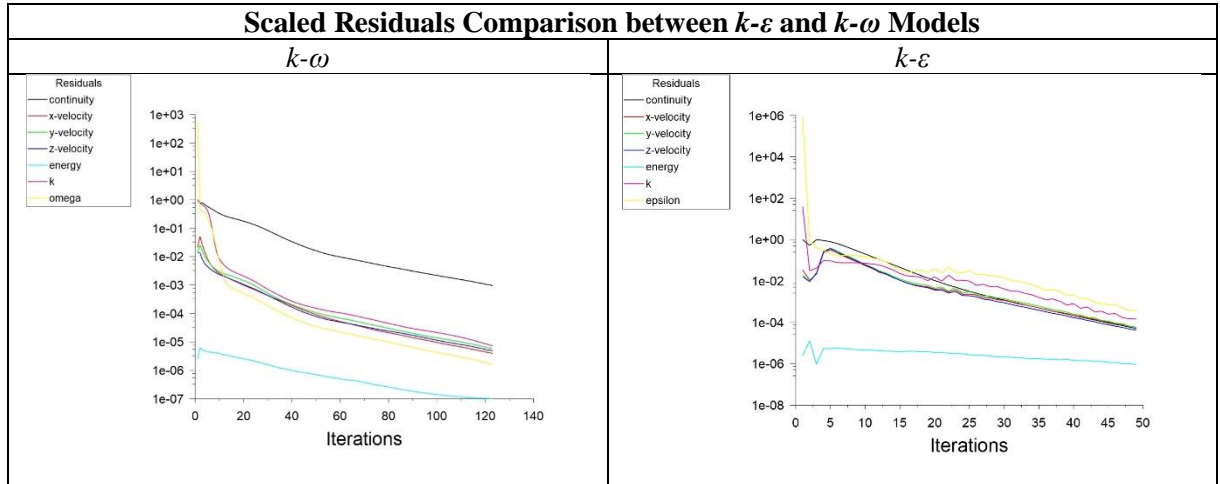

Table 13: Scaled Residual Comparison between k-ε and k-ω Models

The residuals are a function of the error in the solution. For the *k-ε* model, it can be seen in Table 13 that the graph stabilizes by 50 iterations with an RMS residual level close to 1e-03, whereas, in the *k-ω* model it takes 120 iterations for the convergence to stabilize with an RMS residual level close to 1e-06.

The RMS residual level indicated that the k-ε model is loosely converged and the *k-ω* model is tightly converged. For a numerically accurate solution, it is preferable to have lower residual values. Hence, we can say that the *k-ω* model is more efficient.

## 4.2 HEAT TRANSFER RATE OF OUTLET INNER TUBE

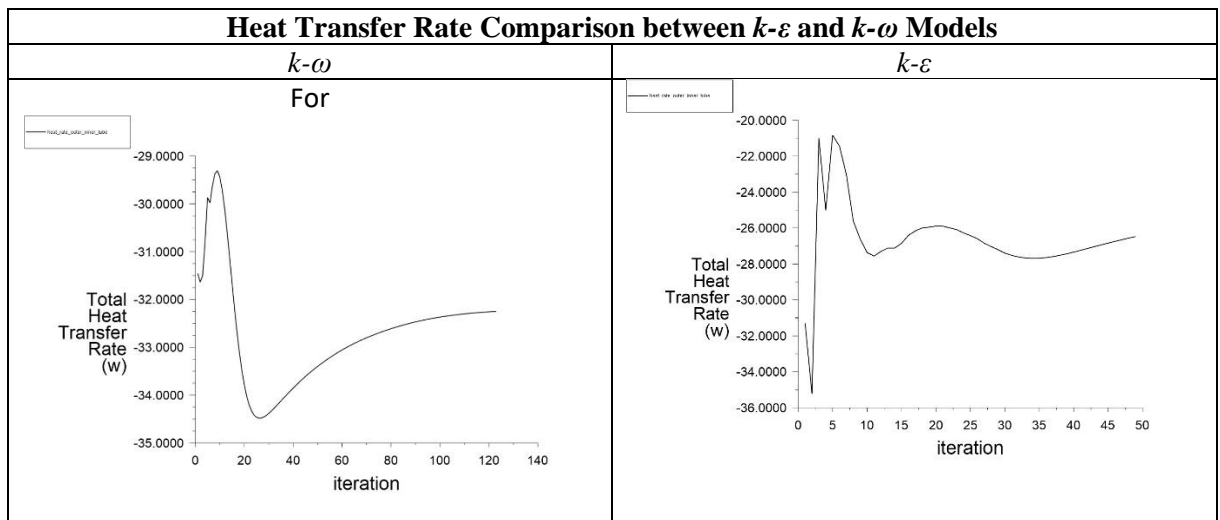

Table 14: Heat Transfer Rate of Outlet Inner Tube Comparison

The total heat transfer rate of the outlet of the inner tube for both models is plotted and compared, shown in Table 14. The total heat transfer rate for the *k-ω* model is marked by an increase in the heat transfer rate followed by a drop. After 25 iterations, the heat transfer rate increases.

In the k-ε model, the total heat transfer rate is marked by sharp drops and rises. The total heat transfer rate is observed at the point of convergence is much higher in the *k-ω* model than in the in the *k-ε* model. Therefore, k- *ω* model is more efficient.

## 4.3 HEAT TRANSFER RATE OF OUTLET OUTER TUBE

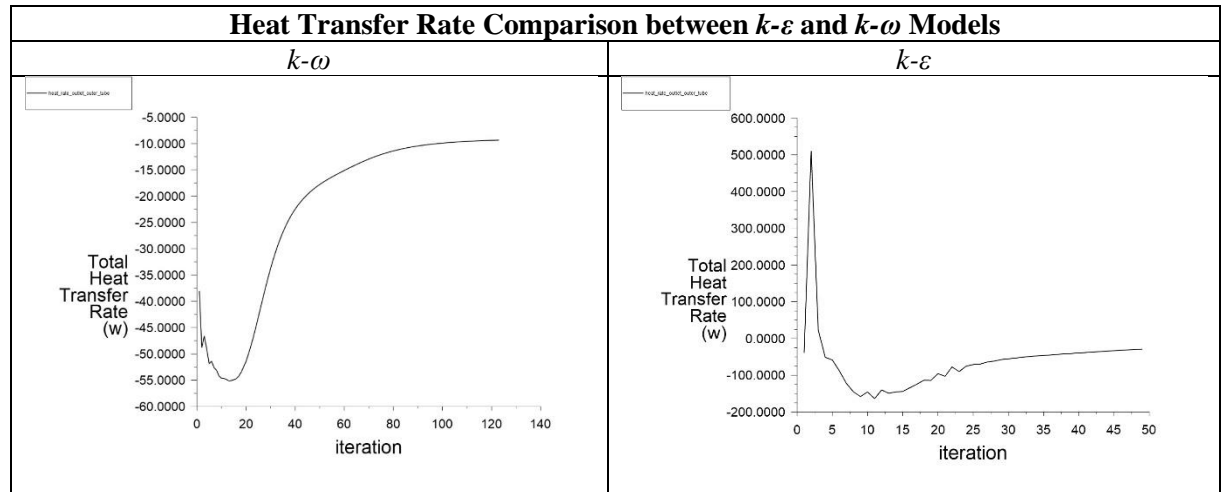

*Table 15: Heat Transfer Rate of Outlet Outer Tube*

Similarly, the total heat transfer rate of the outlet of the outer tube for both models is plotted and compared in Table 15. The total heat transfer rate for the *k-ω* model is marked by a drop, following which the rate increases steadily.

In the *k-ε* model, the total heat transfer rate is marked by a sharp increase followed by a sharp drop in the heat transfer rate, and the curve is not as steady. The total heat transfer rate at the point of convergence is observed to be higher in the in the *k-ω* model than in the *k-ε* model.

## 4.4 TEMPERATURE CONTOURS OF HE

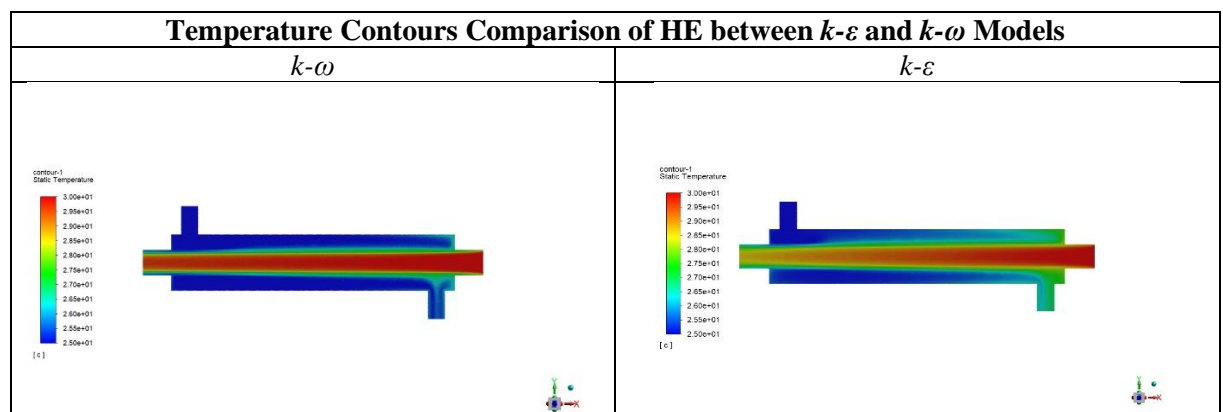

*Table 16: Temperature Contour Comparison*

The temperature contours for *k-ω* and k-ε models are compared. The *k-ω* model is known to provide more accurate solutions in the near wall boundary regions, whereas *k-ε* is best suited for flow away from the wall, i.e., free surface flow region.

The same is proved by the temperature contours shown in Table 16. *k-ω* can be seen to exchange heat better near the walls of the tube in the heat exchanger and *k-ε* has better heat transfer away from the walls of the tube.

## 4.5 PATHLINES OF HE

| Pathline Comparison of HE between *k-ε* and *k-ω* Models | |
|---|---|
| *k-ω* | *k-ε* |
| 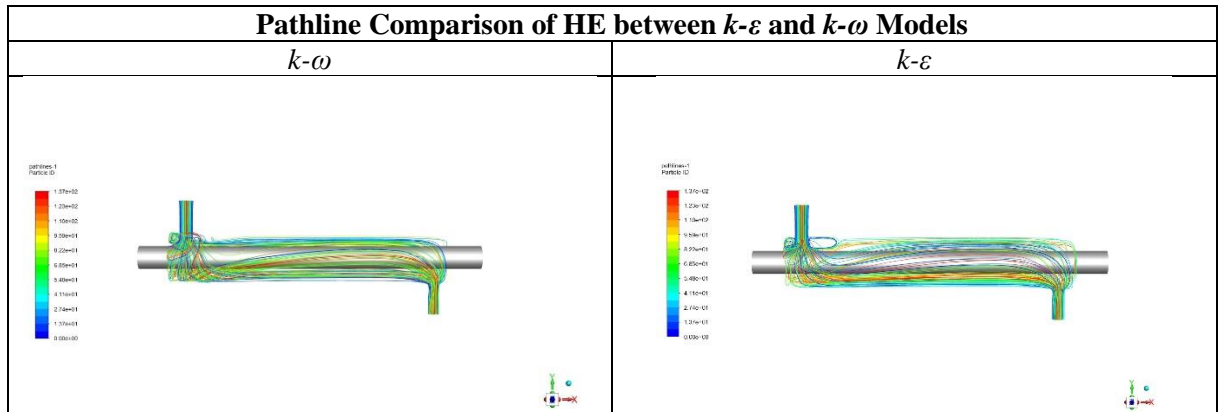 | |

*Table 17: Pathline Comparison*

The pathlines for *k-ω* and *k-ε* model were plotted in Table 17 (observations made from *k-ω* shows the pathlines more saturated than the pathlines from *k-ε*).

# 5 CONCLUSION

Heat transfer is considered a fundamental process in all manufacturing industries. The general function of a heat exchanger is to transfer heat from one fluid to another. The performance and efficiency of heat exchangers are measured by the amount of heat transferred using the smallest heat transfer area and pressure drop. The three-dimensional ANSYS FLUENT CFD model, for shell and tube heat exchanger, was investigated for two two-equation turbulent models, *k-ω* and *k-ε*. The shell and tube heat exchanger has been simulated for a counter-current flow, and the heat transfer rate was obtained for the two models.

The low RMS residual levels calculated showed that the convergence is better in the k-ω model. The differences in the heat transfer rate noted are likely due to the low number of iterations in the k-ε model and a loose convergence. The study of the effect of residuals and iterations on the flow and heat transfer helps explain the selection of an appropriate model.

In the operating conditions used for the simulation, it is observed that *k-ω* model has better heat transfer rate. It is arguable that the heat transfer was not accurate in the k-ε model owing to a lesser number of iterations leading to a higher residual level. Hence, it is safe to conclude that *k-ω* model is best suited for the heat exchanger.

# 6 BIBLIOGRAPHY

1. Heat Exchanger Application General. *Engineers Edge.* [Online] https://www.engineersedge.com/heat_exchanger/heat_exchanger_application.htm.

2. *FUNDAMENTALS OF HEAT EXCHANGERS.* Dr. Osama Mohammed, Elmardi Suleiman Khayal. 2018, INTERNATIONAL JOURNAL OF RESEARCH IN COMPUTER APPLICATIONS AND ROBOTICS , p. 11.

3. What is a heat exchanger ? *Sterling thermal technology.* [Online] https://www.sterlingtt.com/2021/07/14/what-is-a-heat-exchanger/.

4. Kern, D. Q. *Process Heat Transfer.* s.l. : McGraw-Hill , 1983.

5. Robert Serth, Thomas Lestina, Robert Serth. *PROCESS HEAT TRANSFER.* s.l. : Academic Press, 2007.

6. Cao, Eduardo. *Heat Transfer in Process Engineering, 1st Edition.* s.l. : McGraw-Hill Education, 2010.

7. *Numerical Investigation of Shell and Tube Heat Exchanger for Heat Transfer Optimization .* Sahil Suman, Vivek Kumar Chaudhary, Ayush Raj, Kuldeep Rawat. 2017, International Journal of Scientific & Engineering Research, p. 8.

8. Bengt Andersson, Ronnie Andersson, Love Hakansson, Mikael Mortensen, Rahman Sudiyo, Berend van Wachem. *Computational Fluid Dynamics for Engineers.* s.l. : Cambridge, 2012.

9. Malalasekera, H K Versteeg and W. *An Introduction to Computational Fluid Dynamics.* s.l. : Pearson Education Limited, 2007.

10. Ozden, Tari. Shell side CFD analysis of a small shell-and-tube heat exchanger. May 2010.

11. Use of k-epsilon and k-omega Models. *CFD Online.* [Online] https://www.cfd-online.com/Forums/main/75554-use-k-epsilon-k-omega-models.html.